\documentstyle[epsfig]{aipproc}

\begin{document}
\title{Dynamical Stability and Galaxy Evolution in LSB Disk Galaxies}

\author{Chris Mihos,$^*$ Stacy McGaugh,$^{\dagger}$
and Erwin de Blok$^{\ddagger}$}
\address{$^*$Hubble Fellow, Department of Physics and Astronomy, 
	Johns Hopkins University\\
$^{\dagger}$Carnegie Institute of Washington, Department of
        Terrestrial Magnetism\\
$^{\ddagger}$Kapteyn Astronomical Institute, University of Groningen}

\maketitle

\begin{abstract}
We demonstrate that, due to their low surface mass density and large dark 
matter content, LSB disks are quite stable against the growth of global
bar modes.  However, they may be only marginally stable against local disk 
instabilities. We simulate a collision between an LSB and HSB galaxy and 
find that, while the HSB galaxy forms a strong bar, the response of the 
LSB disk is milder, in the form of spiral features and an oval distortion. 
Unlike its HSB counterpart, the LSB disk does not suffer strong inflow of 
gas into the central regions.  The lack of sufficient disk self-gravity to 
amplify dynamical instabilities makes it difficult to explain strong 
interaction-driven starbursts in LSB galaxies without invoking mergers.
\end{abstract}


The lack of companions around low surface brightness (LSB) disk galaxies 
\cite{B93,Tay}
has led to the suggestion that, without the well-established
dynamical trigger provided by interactions, LSB galaxies may
simply evolve passively due to their low surface densities \cite{vdH}, and never
experience any strong star-forming era in their lifetimes. Indeed,
sufficient tidally induced star formation in LSB disks may drive
evolution from LSB to high surface brightness (HSB) galaxies. This has been 
suggested as the cause of the
observed isolation of LSB galaxies: interactions in denser environments
transform them into HSB or HII galaxies or perhaps even destroy them 
entirely.

However, the ability for interactions to trigger evolution and starburst
activity is linked to instabilities in the stellar disk. As LSB disk
galaxies have lower disk mass densities and a greater fraction of dark to 
visible matter than do HSB galaxies\cite{dBM}, the stability of LSB disks -- and their 
response to tidal interactions -- may be quite different than that of 
``normal'' HSB galaxies. In this study, we use analytic stability criteria 
and numerical simulation to investigate the stability of LSB disks in the 
context of galaxy interactions.

\section*{Stability Criteria}

To study disk stability, we use the structural properties of the LSB
disk galaxy UGC 128 and the HSB galaxy NGC 2403, derived by de Blok \& 
McGaugh \cite{dBM} from HI rotation curve decompositions. UGC 128 has
a disk mass density nearly an order of magnitude below that of NGC 2403,
and is more dark matter dominated: the mass-to-light ratio within 6 scale
lengths is $\Upsilon_B=30$ for UGC 128 and $\Upsilon_B=7.4$ for NGC 2403
(see \cite{dBM} for details). The rotation curves for UGC 128 and NGC 2403 
are shown in Figure 1a.

One measure of the susceptibility of galactic disks to global bar
instabilities is the $X_2$ parameter\cite{T81}: 
$X_{m=2} = {{\kappa^2 R}\over{4\pi G \Sigma_d}},$
where $\kappa$ is the epicyclic frequency, $R$ is the radius, and
$\Sigma_d$ is the disk surface density. For flat rotation curves,  disks prove
stable against growing modes if $X_2>3$, while for linearly rising rotation 
curves $X_2>1$ is a sufficient condition for stability. Figure 1b shows $X_2$ as 
a function of scale length for our representative galaxies.  The HSB galaxy 
NGC 2403 is only marginally stable over a large range of radius, while
the LSB galaxy UGC 128 proves stable throughout the disk, due to its lower 
mass surface density. We point out that the rotation curve modeling
assumed maximum disk models; if LSBs are less than maximal disks,
they will be even {\it more} stable.

\begin{figure}[b!] 
\centerline{\epsfig{file=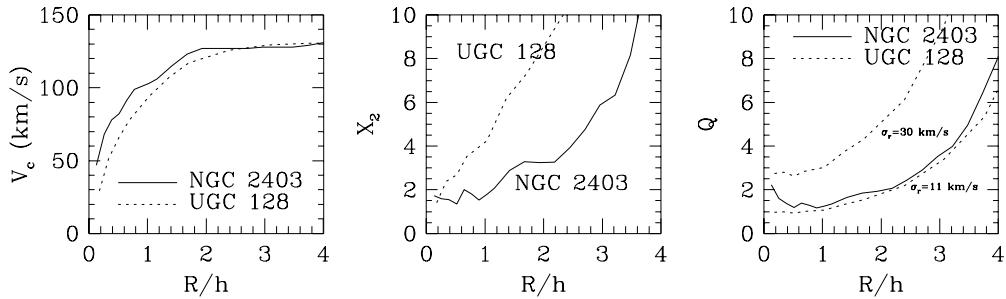,height=1.6176in,width=5.3in}}
\vspace{10pt}
\caption{Left: Rotation curves of NGC 2403 (HSB) and UGC 128 (LSB),
as a function of disk scale length ($R/h$).  Middle:
$X_2$ stability parameter. Right: Toomre $Q$ parameter. The two curves for
UGC 128 reflect two choices for $\sigma_r$.}
\label{fig1}
\end{figure}

If LSB disks are stable against the growth of global instabilities in the
disks, are they also stable against {\it local} instabilities? The growth
of local axisymmetric instabilities is measured by the Toomre $Q$ parameter
\cite{T64}: $Q = { {\sigma_r \kappa} \over {3.36 G \Sigma_d} }$,
where $\sigma_r$ is the radial velocity dispersion of the disk stars.
Lacking information on $\sigma_r$ in LSB disks, we use two alternatives:
1) that $\sigma_r$ is like that in the Milky Way ($\sim$ 30 km s$^{-1}$) or
2) that $\sigma_r^2 \sim \Sigma_d$ (so that $\sigma_r \sim$ 11 km s$^{-1}$).
Figure 1c shows Q in each disk; if velocity dispersion drops with surface
density as might be expected from energy arguments, LSB and HSB disks may have similar {\it local} stability properties, 
such that local instabilities might grow in LSB disks where global modes 
cannot.

\section*{Numerical Models}

To examine how LSB disks will respond to a close interaction, we simulate a
grazing encounter between an LSB galaxy and an HSB companion. We choose a
prograde, parabolic orbit with a perigalactic separation of $R_p=10$ disk
scale lengths. 

Rather than build galaxy models which differ in a number of structural
parameters, we focus on variations in disk surface density to define
the difference between HSB and LSB disk galaxies. We construct two
model galaxies with disk surface densities which differ by a factor
of eight, similar to the difference between NGC 2403 and UGC 128.
The dark halos have identical mass distributions (as a function of $R/h$)
in both galaxies, resulting in our LSB being very dark matter dominated.
We initialize velocities in {\it both} galaxy 
disks such that Q=1.5, implying lower velocity dispersion
in the LSB disk; the simulation is thus a conservative test of
LSB stability. In models which include gas, the gas comprises 10\% of
the total disk mass in each galaxy.

Figure 2 shows the evolution of the HSB and LSB disks in the stellar
dynamical interaction model. Both galaxies respond strongly during the
close passage (at T=24). In the HSB disk, the self-gravity
of the disk amplifies the perturbation such that by T=44 the galaxy has
developed a very strong bar. By contrast, the LSB disk displays
a persistent oval distortion and long-lived spiral arms in the disk.
Without adequate disk self-gravity 
no strong bar develops in the LSB disk. Figure 3a shows the strength
of the $m=2$ mode in the inner half mass of each disk. The peak
strength is more than twice that of the LSB disk, and declines at
late time, probably due to disk heating by the bar. We emphasize that
the $m=2$ mode is not only different in strength between the disks,
but also in character: the HSB sports a strong bar, while the LSB displays
a milder oval distortion. The bar in the HSB galaxy drives strong inflow
(Figure 3b): the gas surface density in the center of the HSB disk has
risen significantly by T=36.\footnote{At this point, the gas was ``switched
off'' in the HSB to save computational expense; however, inflow was
ongoing, and the final gas density at the center of the HSB would be
even higher than shown here.} By contrast, the relatively weak 
response of the LSB disk results in very little change in the gas mass 
distribution in the disk, even much later after the encounter at T=72
(Figure 3c).

\begin{figure}[b!] 
\centerline{\epsfig{file=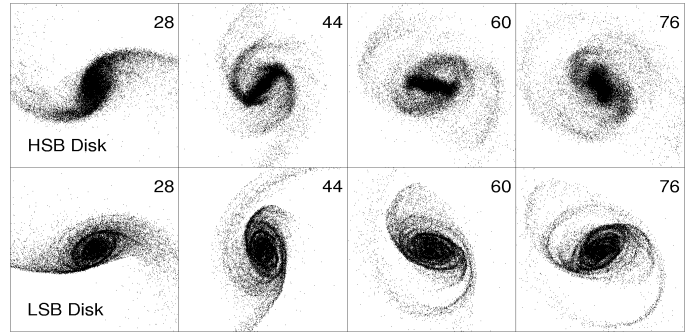,height=2in,width=3.66in}}
\vspace{10pt}
\caption{Post collision disk evolution. Top: HSB disk. Bottom:
LSB disk. Each frame is 10 scale lengths on a side, and time is given
in the upper right. One rotation period is approximately 13 time units.}
\label{fig2}
\end{figure}

\begin{figure}[b!] 
\centerline{\epsfig{file=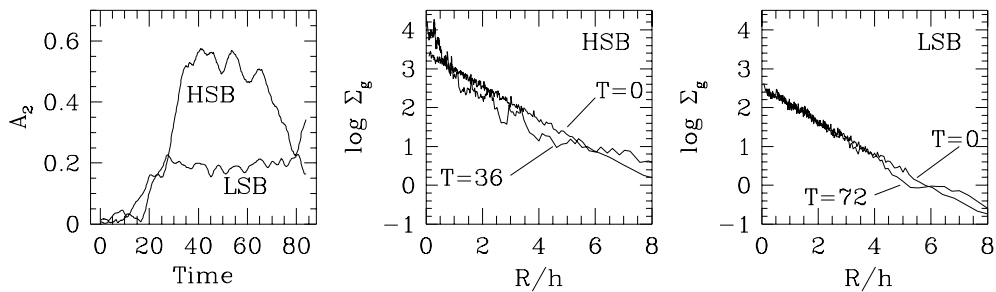,height=1.726in,width=5.3in}}
\vspace{10pt}
\caption{Right: Growth of m=2 modes in stellar-dynamical simulation.
Middle: Gas mass profile in HSB disk in stellar+hydro simulation.
Right: Gas mass profile in LSB disk in stellar+hydro simulation.}
\label{fig3}
\end{figure}

\section*{LSBs and Galaxy Evolution}

Both analytic arguments and numerical simulation indicate that, despite
their seemingly fragile nature, LSB disks are quite stable, and resistant
against the growth of bars and bar-driven inflows. These results present
a problem for the otherwise appealing notion that interacting LSB dwarfs 
are the progenitors of HII galaxies experiencing central starbursts 
\cite{Tay}. Even the relatively close, strong interaction we have presented 
will not result in a strong central starburst, nor will it drive strong
structural evolution in the galaxy; in order to provoke a violent enough 
response in the LSB disk, a bona-fide merger may be necessary.

\end{document}